**Niche inheritance: a cooperative pathway to enhance cancer cell fitness though ecosystem engineering**

Kimberline R. Yang[4]

Steven Mooney[1]

Jelani C. Zarif[1]

Donald S. Coffey[1,2,3]

Russell S. Taichman[5]

Kenneth J. Pienta[1,2,3,4]

Authors' Affiliations: [1]The James Buchanan Brady Urological Institute and Departments of Urology, [2]Oncology, [3]Pharmacology and Molecular Sciences, [4]Cellular and Molecular Medicine Program, Johns Hopkins School of Medicine, Baltimore, Maryland; [5]Department of Periodontics and Oral Medicine, University of Michigan School of Dentistry, Ann Arbor, Michigan

Corresponding Author: Kenneth J. Pienta, The James Buchanan Brady Urological Institute and Department of Urology, Departments of Oncology and Pharmacology and Molecular Sciences, The Johns Hopkins School of Medicine, 600 N Wolfe St, Baltimore, MD 21287.
E-mail: kpienta1@jhmi.edu

**Key words:** niche construction, diaspora, metastasis, genetic instability, tumor cell heterogeneity, dispersal filters

**Grant Support**
This work is financially supported by the NCI (grant no. U54CA143803 to K.J. Pienta, D.S. Coffey; grant nos. CA163124 and CA093900 to K.J. Pienta, R.S. Taichman; and grant no. CA143055 to K.J. Pienta).

**Abstract:**

Cancer cells can be described as an invasive species that is able to establish itself in a new environment. The concept of niche construction can be utilized to describe the process by which cancer cells terraform their environment, thereby engineering an ecosystem that promotes the genetic fitness of the species.  Ecological dispersion theory can then be utilized to describe and model the steps and barriers involved in a successful diaspora as the cancer cells leave the original host organ and migrate to new host organs to successfully establish a new metastatic community. These ecological concepts can be further utilized to define new diagnostic and therapeutic areas for lethal cancers.

## Introduction

The cooperative construction of a new tumor niche through ecological engineering is a keystone event for the formation and function of a cancerous lesion.   The evolving niche changes the ordered organ microenvironment into a disordered malignant microenvironment that in turn changes the genotypes and phenotypes of both cancer and host cells. This new heterogeneous environment is built in a cooperative manner between cancer and host cells, induces a high rate of tumor cell heterogeneity.  Natural selection and therapeutic selection ensue and selected cancer cells survive to continue the process locally or through a diaspora to a distant site.

## Cancer and niche construction

A niche in ecology refers to both the place a species lives as well as the role it plays in its habitat, including the dynamic flow of information and energy around it (Grinnell, 1917; Hutchinson, 1957; Elton, 2001).  It includes how an individual organism, or the population of its species in that ecosystem, utilize and respond to resources, the abiotic environment it interacts with, and the stresses caused by competitors and environmental changes.  The sciences of ecology, evolution, population biology, and sociology have created many paradigms that can be utilized to better understand cancer and the processes of tumorigenesis and metastasis (Chen and Pienta, 2011; Camacho and Pienta, 2012; Pienta et al., 2013; Akiptis et al., 2013; Scott et al., 2013, 2014).  Niche construction theory integrates ecosystem ecology theory and evolutionary dynamics to explain the interplay between a species, its environment and genetic drift (Odling-Smee et al., 2003, 2013; Erwin, 2008; Kylafis and Loreau, 2008; Krakauer et al, 2009; Post and Palkovacs, 2009; Loreau, 2010; Van Dyken and Wade, 2012). Niche construction is the process whereby organisms, through their metabolic activities and other behaviors, modify their own and/or each other's niches (Odling-Smee et al., 2003, 2013).  As a consequence of these behaviors, niche construction may result in changes in one or more natural selection pressures in the external environment of their own or others populations.  Species that construct niches may also be referred to as ecosystem engineers (Jones et al., 1994, 1997; Badano et al., 2006).

## Cancer Cells as Ecological Engineers

Ecosystem engineers construct and modify their niche (Jones et al., 1994, 1997; Badano et al., 2006).  Allogenic engineers modify their environment by mechanically changing their environment ( e.g., beavers). Autogenic engineers modify their environment by changing themselves over time (e.g., trees as they grow) (Jones et al., 1994, 1997; Badano et al., 2006).   Many invasive species funtion as ecosystem engineers as they change the ecosystem around them as they construct a niche that is favorable to their own survival (Hickman et al, 2010; Chen and Pienta, 2011).  Cancer cells function as both allogenic and autogenic engineers (Figure 1).  As allogenic engineers, for example, they secrete matrix metalloproteinases that physically alter their environment (Hanahan and Weinberg, 2011). The secretion of vascular endothelial growth factor (VEGF) attracts the formation of new vasculature to the local tumor ecosystem (Wey et al., 2004; Hanahan and Weinberg, 2011; Catalano et al., 2013; Burkholder et al., 2014). As autogenic engineers, tumors grow in size, changing the architecture, pH, and interstitial pressure of the organ host ecosystem in which they live (Jain, 2012; Barar and Omidi, 2013; Stylianopoulos and Jain, 2013). This fundamentally changes the growth patterns of host cell species as well as changes the flow of nutrients and information in the forms of cytokines, chemokines, hormones and exosomes as they traffic through the ecosystem (Jain, 2012; Barar and Omidi, 2013; Stylianopoulos and Jain, 2013).

Niche construction by an invasive species fundamentally changes the ecosystem in which it establishes. Initially, cancer cells, even when they arise in a primary organ site, act as an invasive species. Odling-Smee theorized that niche construction can cause ecological inheritance (Odling-Smee et al., 2003, 2013). Ecological inheritance is the inheritance, via an external environment, of one or more natural selection pressures previously modified by niche-constructing organisms (Odling-Smee et al., 2003, 2013). The concept of ecological inheritance depends on a species leaving the altered niche to their offspring, i.e., the next generation of the species is born into the engineered environment. This engineered environment can then speed the process of the selection of genetic factors that increases a species' chances of survival. Tumor cell heterogeneity is a well-known concept in cancer biology and is generally attributed to intrinsic genetic instability (Pienta et al., 1989; Hunter, 2006; Hanahan and Weinberg, 2011; Klein, 2013). The concept of ecological inheritance suggests that the production of tumor cell heterogeneity may be increased through niche construction/ecological engineering (Figure 1). Given these findings, it is possible that this is a plausible concept (Figure 2). For example, does the fact that cancer cells create a hypoxic, nutrient – low environment lead to increased clonal heterogeneity or less as only a few adaptive clones survive? In the case of cancer, ecological inheritance of the malignant niche appears to promote the biodiversity of the cancer species (tumor cell heterogeneity), ultimately resulting in the development, selection, and survival of lethal clones.

A fundamental difference between ecological engineers in nature and cancer cells appears to be the health and longevity of the niche that the species constructs. A good example of this is when a beaver creates a pond that supports life and is passed onto its offspring. The beaver creates a beautiful pond, not a stagnant swamp. The human mind automatically assumes and looks for the "healthy" ecosystem. Nothing about a cancer microenvironment looks healthy to us. As cancer cells divide, they outstrip their blood supply, creating a nutrient poor, poorly oxygenated, acidic stagnant swamp rather than a healthy pond. The cancer swamp hardly seems conducive to growth and yet, this is exactly the environment the cancer cells may need to accelerate the generation of adaptive clones that have the ability to metastasize. As a secondary consequence, an environment is created that destroys the niches of normal host cells with resultant organ destruction, i.e., an ecological spillover. Even the beaver drives out normal species (e.g trees living in the upstream drainage) while creating new habitats for non-beaver species (ducks, fish). In much the same way, cancer changes the host cells present in the organ ecosystem, destroying (e.g., epithelial cells) and attracting others (e.g., tumor associated macrophages).

The concepts of ecological engineering and niche construction may have diagnostic and therapeutic implications. Diagnostic tools that detect areas of loss of tissue metabolic homeostasis could potentially lead to earlier cancer detection. Areas of hypoxia or low pH could signify a growing collection of tumor cells. It is possible that metastatic cancer would occur at a much slower rate if cells were not forced to adapt as they are subjected to the stresses of the developing cancer stagnant swamp. Agents that block this adaptation could be developed – a prime target of pharmacological inhibition is HIF-1α (Semenza, 2012; Chaturyedi, 2013). HIF-1α mediates many of the stress response pathways that are the result of hypoxia. The strategic trick would be to use a specific pharmacological inhibitor of HIF-1α early in the process of niche construction at the primary and/or metastatic sites. Similarly, targeting other stress response proteins by repurposing inhibitors to use them early in targeting cancer niche construction events may be fruitful.

**Cancer Metastasis as a Form of Ecological Dispersal**

Once a lethal cancer successfully establishes a niche in the primary organ, it invariably metastasizes (Hanahan and Weinberg, 2011; Klein, 2013; Scott et al., 2013; Lavi et al., 2014). We have utilized the social science concept of diaspora to describe metastasis in terms of the traits a species must have to successfully leave the original host organ and migrate/disperse to new host organs, and then successfully establish a new community (Figure 3)(Pienta et al., 2013). For a species to successfully disperse, it must travel to a new area, tolerate conditions of a new habitat and reproduce (Figure 4). Ecologists have defined types of dispersal events, including diffusion and jump dispersal (Suarez et al., 2001; De Valpine et al, 2008) (Table 1). Diffusion is the slow dispersal of individuals spreading out from the margins of the species' range. This is accomplished over generations and is dependent on multiple factors, including food supply and successful population growth. For example, house sparrows were introduced into North America (jump dispersal event) when birds from England were released in New York City in 1852 and then by diffusion dispersal have spread from Central America to northern Canada. (Johnston et al., 1973; Healy et al, 2009). This concept is analogous to cancer cells at a primary tumor site simply growing in number, resulting in a larger tumor over time, which clearly takes multiple generations of clonal growth and establishment of new blood supplies to allow delivery of oxygen and nutrients. Jump dispersal is a long-distance dispersal over inhospitable terrain accomplished during a relatively short period. It occurs infrequently, but results in the presence of a species in distinct geographical locations, e.g., movement of birds between islands. Cancer cell metastasis through the blood stream is an example of jump dispersal as tumor cells leave the primary organ and travel to establish at distant organs (Chen and Pienta, 2011; Hanahan and Weinberg, 2011; Semenza, 2012; Klein, 2013; Pienta et al., 2013). Just as some species are better at jump dispersal than others (e.g. better suited to survive transport by water or wind), different cancer cells are better suited to survive the jump through the blood stream (Charpentier and Martin, 2013; Lianidou et al., 2013; Lowes and Allan, 2014; Tinhofer et al., 2014). Cells that have undergone EMT or cancer stem cells appear to survive in the circulation better than those with an epithelial phenotype (Charpentier and Martin, 2013).

Barriers to migration include abiotic and biotic features that preclude successful dispersal (Figure 4). These barriers can also be considered "filters" that prevent movement of a species from one place to another, or in the case of cancer, from a primary to metastatic site or, from one primary metastatic site to a secondary metastatic site. In ecology, dispersal barriers or filters are defined as "abiotic or biotic [factors] that restrict movement of genes or individuals from one place to another" (Boulangeat et al., 2012). Multiple organisms such as whales face predation barriers during their migration to warmer waters to breed. Likewise, cancer cells face physical and ecological barriers during their migration to distant organs, analogous to dispersal filters, which include unfavorable environmental conditions in the blood circulation and encounters with the host immune system.

The tight vascular junctions of the endothelial vessels serve as one of the early dispersal filter for potentially successful metastatic cancer cells (Kim et al., 2009; Nguyen et al., 2009; Comen and Norton, 2012) (Figure 4). After successful intravasation, the turbulent bloodstream itself serves as another early dispersal barriers of metastasis since less than 1% of circulating tumor cells survive (Fidler, 1973). Upon entry into the bloodstream, these circulating tumor cells face a foreign and rather harsh environment where they are susceptible to anoikis, a form of programmed cell death triggered by detachment from the extracellular matrix by cells that are normally anchorage-dependent (Weiss et al., 1981; Faraji and Eissenberg, 2013; Ramakrishna and Rostomily, 2013). Unlike red blood cells, tumor cells are not able to withstand the shear force of the rapid blood flow (Faraji and Eissenberg, 2013). In addition, tumor cells have three to four times wider diameter than capillaries and some appear to be more rigid or more prone to cluster, which can trap them in the narrow vessels and cause them to die in circulation before reaching their preferred secondary site (Fidler, 1970; Faraji and Eissenberg, 2013; Plaks et al., 2013).

The concept of an artificial dispersal filter for diagnosis and therapy in the form of an ecological trap is an intriguing one. Ecologic traps are poor-quality habitats that are highly attractive to wildlife species based on environmental cues that typically signify a high-quality habitat (Shiozawa et al., 2011; Li and Mooney, 2013; Pienta et al., 2013, Robertson et al., 2013; Van der Sanden et al., 2013). A prototypical example is a mosquito being attracted to a bright light and then dying from the heat. An indwelling filter in the blood stream, infused with a chemoattractant such as stromal derived factor -1 (SDF-1), could catch circulating tumor cells (Shiozawa et al., 2011).

In addition, circulating tumor cells interact with different cell types, many of which are the host immune cells that can recognize and eliminate cancer cells (Tarhini et al, 2014) (Figure 4). Immune cells are constantly circulating the bloodstream and monitoring for any foreign species. Unlike bacteria, viruses, or parasites, cancer cells are not foreign to the host. However, because of aberrant changes in their genetic makeup and cell biology, they may express antigens distinct from normal host cells (54). Expression of tumor antigens can be recognized by circulating leukocytes such as natural killer cells and $CD8^+$ T cells that can recognize tumor antigens presented by MHC molecules and trigger cytokine release to recruit macrophages, eosinophils and mast cells as well as trigger lysis or apoptosis of the tumor cell (Zitvogel et al., 2008). The greater number of immune cells in circulation compared to the number in the primary tumor allows immune cells to effectively eliminate disseminated tumor cells (DTCs) (Knutson and Disis, 2005; Wan et al., 2013).

Disseminated cancer cells that are able to surpass these dispersal filters are able to successfully leave their primary tumor sites to reach their target organs where they may undergo self-renewal to establish a new colony. There is still a barrier, however between reaching their target and successfully self-seeding there. These sets of barriers are referred to as niche filters (Maire et al., 2012; Thuiller et al., 2013). In ecology, niche filters "select for species that can establish and maintain positive population growth under the given environmental conditions" (Maire et al., 2012; Thuiller et al., 2013). These selective pressures include species fitness, abiotic environmental conditions and biotic inter-species competition (Maire et al., 2012; Thuiller et al., 2013). In order to establish a new niche in the secondary target organ, cancer cells must overcome niche filters such as "soil" quality, host cell occupancy and the immune system.

In 1889, Stephen Paget highlighted the importance of the soil as well as the organ microenvironment or niches for metastatic colonization in his seed and soil hypothesis (Paget, 1889; Matho and Stenninger, 2012). Many types of cancer metastases show organ-specific dissemination, such as breast and prostate cancer to the bone marrow (Nguyen et al., 2009). The seeding/colonization potential of metastatic cancer cells depends largely on specific molecular interactions between the cancer cells and the host microenvironment of the metastatic site. The soil quality is defined by how receptive a particular target organ is to DTCs. It is determined by factors in the tumor microenvironment that facilitate the successful survival and colonization of disseminated cancer cells. These factors include extracellular matrix (ECM) components and basement membranes, stromal cell types, chemokines, cytokines, and hormones, reactive oxygen species, the availability of nutrients and oxygen, and presence of immune system cells (Gupta and Massagué, 2006; Steeg, 2006; Oskarsson et al., 2014).

ECM components are the first physical barrier for DTC invasion of the secondary site (Figure 4). In order for DTCs to successfully land and colonize distant organ sites, appropriate interactions with specific adhesion and signaling molecules are required.

These signals are crucial for proliferative signaling cascades within the cells. For example, breast cancer cells require binding interactions with ECM components such as collagen I and fibronectin in the lung parenchyma via $β_1$-containing integrins for FAK-mediated proliferation in the lung (Shibue and Weinberg, 2009; Wan et al., 2013). Cells that lack these pro-proliferative interactions undergo apoptosis and therefore are unable to survive at the secondary organ site.

Stromal cell types also determine the viability of DTCs at target organ sites. Both breast and prostate cancer metastasis localize to the bone marrow. The bone marrow niche houses a large number of stromal cells such as osteoblasts, endothelial cells, adipocytes, mesenchymal stem cells, and CXCL12-abundant reticular cells (Pedersen et al., 2012). Osteoblasts secrete the cytokine SDF-1 that interacts with CXCR4 or CXCR7 receptors on prostate cancer cells to stimulate the invasion and homing to the bone marrow. Disrupting the SDF-1/CXCR4 pathway by either depleting SDF-1 or blocking CXCR4 or CXCR7 receptors disrupts the ability of prostate cancer cells to colonize the hematopoeitic stem cell niche (Pedersen et al., 2012). In addition, competition with stromal-derived growth-suppressive signals such as bone morphogenic protein (BMP) in the lung parenchyma can hinder colonization (Wan et al., 2013).

The action of osteoblasts also highlights the importance of soluble factors such as chemokines, cytokines, and hormones or growth factors in influencing the tumor microenvironment. Tumor cells can also secrete factors such as TGFβ to remodel the target organ to be more receptive to DTC homing or to prime themselves for organ infiltration (Wan et al., 2013). However, many organ microenvironments are non-receptive to tumor cell signals or express signals incompatible with tumor cell survival and therefore pose a threat to the viability of DTC seeds (Nguyen et al., 2009).

Reactive oxygen species (ROS) are another important factor in the quality of the secondary tumor microenvironment. ROS including free radicals and peroxides are natural byproducts of aerobic metabolism in normal cells. In the absence of tight regulation, excess ROS can induce oxidative stress, DNA damage and DNA mutations to initiate tumorigenesis (Waris et al, 2006; Nishikawa, 2008; Sreevalsan and Safe, 2013). However, over-accumulation of ROS can also activate apoptotic pathways and suppress proliferative signals that threaten the survival of cancer cells (Sreevalsan and Safe, 2013).

DTCs that survive upon encounter with the target organ require sufficient nutrients and oxygen to initiate seeding. For example, cancer cells require angiogenesis for growth and expansion of the tumor via the diffusion of nutrients from blood vessels. Cancer cells that cannot activate the angiogenic "switch" upon arrival at the target organ or that are far from capillaries are unable to form viable colonies and undergo apoptosis or dormancy (Folkman, 1971; Zetter, 1998). At the same time, some but perhaps not all DTCs which arrive in a target organ are either induced to become dormant or may initially lack the machinery for growth in a diaspora setting. Additional genetic lesions may be required prior to the emergence of metastatic outgrowths or may need to terraforming their new environment to establish conditions suitable for growth.

In addition to coping with the compatibility of the soil niche, DTCs must compete with the native host cells for available nutrients and survival signals. The ecosystem of the tumor microenvironment is characterized by the dynamic interactions between the organisms, which in this case are the cancer cells and host cells. Similar to ecological communities, these organisms compete with each other to survive in an environment with limited resources (Pienta et al., 2008). While the metastatic site is completely occupied with native cell populations, only a minority of DTCs survives the dispersal filters and barriers upon initial arrival. Therefore, based on population size, DTCs are already at a disadvantage to the host cells (Gatenby, 1991). Furthermore, competitive interactions between the two cell populations can activate tumor suppressive mechanisms to favor wildtype cells. Surrounding host cells can sense the presence of aberrant cells and eliminate them by extrusion from the tissue epithelium and induction of growth arrest, differentiation, engulfment, and apoptosis. Other mechanisms include secretion of cytotoxic soluble ligands such as IL-25 and secretion of tumor suppressive microRNAs such as miR143 that inhibit tumor proliferation (Wagstaff et al., 2013). While DTCs compete with wildtype host cells to survive in their new niche, they are also highly susceptible to the resident immune cells at the metastatic site.

A subset of immune cells that traffic to the metastatic microenvironment are called tumor infiltrating lymphocytes (TILs) (Slaerno et al., 2014). These include macrophages, dendritic cells, natural killer cells, B cells and effector T cells (Fridman et al., 2012). $CD8^+$ cytotoxic T cells have been largely implicated in antitumor immunity. Similar to their circulating counterparts, $CD8^+$ T cell infiltrates recognize tumor peptide antigens, present them to MHC class I molecules and release cytokines to induce the killing of tumor cells (Yu and Fu, 2006; Mitchell et al., 2014).

The niche filter barrier is another therapeutic target. Circulating tumor cells appear to intravasate into a target organ and then undergo undergo a period of dormancy before starting the process of niche construction and naturalization that results in a clinical metastasis (Gupta and Massagué, 2006; Steeg, 2006; Atkipis et al, 2013; Pienta et al., 2013; Oskarsson et al., 2014, Scott et al., 2013, 2014). Mobilization of these cells prior to their proliferation could lead to their destruction and an interruption of the diaspora process. Shiozawa and colleagues demonstrated the ability of AMD3100, an inhibitor to the receptor for SDF-1, CXCR4, to mobilize prostate cancer cells out of the bone marrow and into the circulation where they could be destroyed (Wang et al., 2006; Shiozawa et al., 2011).

For DTCs, successful colonization of the target organ remains a challenge because of these niche filters. Their survival in the foreign microenvironment is determined largely by their interactions with new cell types and cell substrates that induce multiple molecular mechanisms to combat the presence and colonization of the mutant cells. Yet

DTCs have evolved to become highly resistant against the host response. Metastasis still remains the cause of 90% of cancer-related deaths (Loberg et al., 2007).

**Conclusions: The Cancer Species Niche Construction Paradox**

In nature, many invasive species act as a ecological engineers to create a niche that is conducive to its survival. From an ecological perspective, cancer appears to not make sense because it does not create an ecosystem that achieves equilibrium or a steady-state that allows it to survive as a species – it does not construct a stable niche. But it does engineer a niche that allows it to perpetuate itself and spread (Figure 4). Since there is no negative feedback or control, it ultimately causes organ destruction and the death of the host and itself. From an evolutionary standpoint then, cancer does not appear to be successful. This all depends on perspective.

In ecological and evolutionary terms, cancer is the prototypical 'successful' invasive species when looked at in terms of generation and time scale. It lives for thousands of generations and constructs a primary niche that forces it to acquire added qualities that then allow it to spread and invade new environments. Often it is only stopped by the death of the host biosphere. All species in nature live within the earth's biosphere and species that survive and propagate within it are considered successful – but this will only be true while the earth remains healthy.

Table 1.  Types of dispersal events in earth ecology and cancer ecology.

| Types of Dispersal Events | Earth Ecology | Cancer ecology |
|---|---|---|
| **Jump dispersal:**  long distance dispersal accomplished during a relatively short period of time (occurs infrequently but explains species in different sites) | Movement of species with wind, or carried by artificial means (movement of sparrows from England to North America) | Movement of cancer cells through the bloodstream. |
| **Diffusion**: Slow dispersal of individuals spreading out from the margins of the species' range (accomplished over generations | Movement of species as they reproduce and move to nearby favorable environments (sparrows in North America) | Growth of a primary tumor or a cancer at a metastatic site. |

**Figure Legends:**

**Figure 1. Cancer cells as ecological engineers.** Ecosystem engineers construct and modify their niche to create environmental conditions that favor their survival. Cancer cells, for example, function as engineers as they secrete matrix metalloproteinases that physically alter their environment, attract the formation of new vasculature, change the architecture, pH, and interstitial pressure of the organ host ecosystem in which they live. This fundamentally changes the growth patterns of host cell species as well as changes the flow of nutrients and information in the forms of cytokines, chemokines, hormones and exosomes as they traffic through the ecosystem. Tumor cell heterogeneity is promoted through inherent genetic instability as well as the ecological inheritance through adaptive selection.

**Figure 2. Modeling ecological inheritance.** Niche construction by a species fundamentally changes the ecosystem in which it establishes. The theory of ecological inheritance describes the inheritance, via an external environment, of one or more natural selection pressures previously modified by the ecological engineer species. Ecological inheritance depends on a niche existing across multiple generations of a species, i.e., the next generation of the species is born into the engineered environment. This engineered environment can then speed the process of the selection of genetic and epigenetic factors that increase a species' chances of survival. Gene pool 1 reflects the amount of tumor cell heterogeneity that is a result of the intrinsic genetic instability of cancer cells. Gene pool 2 reflects the increased amount and rate of genetic instability as a result of the malignant niche environment created by the ecological engineering of the cancer cells. Ultimately, this results in increased fitness of the species as cancer cell clones are generated that have the attributes necessary for survival and metastasis.

**Figure 3. The cancer diaspora.** The diaspora paradigm takes into account and models several variables in the metastatic cascade. A diaspora is started by unfavorable conditions in a homeland, leading to the voluntary or forced eviction of a population. The nutrient poor and hypoxic environment of the evolving primary tumor microenvironment reflects this. The diaspora concept also accounts for the fitness of individual cancer cell migrants and migrant populations. Since diaspora communities remain in contact with their homeland, it also describes and models the bidirectional movement of cancer and host cells between cancer sites (including between primary and metastases as well as between metastases). By describing the receptivity of the new hostland for the arriving migrants, the diaspora also models the quality of the target microenvironments to establish metastatic sites (adapted from 4).

**Figure 4. Cancer Metastasis as a Form of Ecological Dispersal.** Once a cancer successfully establishes a niche in the primary organ, it invariably metastasizes. Disseminated cancer cells use the blood stream to undergo jump dispersal and if they are able to surpass dispersal and niche filters they can act as an invasive species and establish a foothold in distant sites. Eventually they may proliferate and act as ecological engineers to form a new niche in the target organ.

References:

Aktipis C.A., Boddy A.M., Gatenby R.A., Brown J.S., Maley C.C. 2013. Life history trade-offs in cancer evolution. Nat Rev Cancer (12):883-92.

Badano, E.I., Cavieres, Lohengrin A. 2006. Ecosystem engineering across ecosystems: do engineer species sharing common features have generalized or idiosyncratic effects on species diversity? Journal of Biogeography (33,2):304-313

Barar J., Omidi Y. 2013. Dysregulated pH in Tumor Microenvironment Checkmates Cancer Therapy. Bioimpacts (4):149-162.

Boulangeat I., Gravel D., Thuiller W. 2012. Accounting for dispersal and biotic interactions to disentangle the drivers of species distributions and their abundances Ecology Letters (15):584–593

Burkholder B., Huang R.Y., Burgess R., Luo S., Jones V.S., Zhang W., Lv Z.Q., Gao C.Y., Wang B.L., Zhang Y.M., Huang R.P. 2014. Tumor-induced perturbations of cytokines and immune cell networks. Biochim Biophys Acta 1845 (2):182-201

Camacho D.F., Pienta K.J. 2012. Disrupting the networks of cancer. Clin Cancer Res 18(10):2801-8.

Catalano V., Turdo A., Di Franco S., Dieli F., Todaro M., Stassi G. 2013. Tumor and its microenvironment: a synergistic interplay. Semin Cancer Biol (6 Pt B):522-32.

Charpentier M., Martin S. 2013. Interplay of Stem Cell Characteristics, EMT, and Microtentacles in Circulating Breast Tumor Cells. Cancers (Basel)5(4):1545-65.

Chaturvedi P., Gilkes D.M., Wong C.C., Kshitiz, Luo W., Zhang H., Wei H., et al. 2013. Hypoxia-inducible factor-dependent breast cancer-mesenchymal stem cell bidirectional signaling promotes metastasis. J Clin Invest (123):189–205.

Chen K.W., Pienta K.J. 2011. Modeling invasion of metastasizing cancer cells to bone marrow utilizing ecological principles. Theor Biol Med Model 8:36

Comen E, Norton L. 2012. Self-seeding in cancer. Recent Results Cancer Res (195): 13-23

De Valpine P., Cuddington K., Hoopes M.F., Lockwood J.L. 2008. Is spread of invasive species regulated? Using ecological theory to interpret statistical analysis. Ecology (9):2377-83.

Elton, C. S. 2001. Animal Ecology. University of Chicago Press ISBN 0-226-20639-4.

Erwin D. H. 2008. Macroevolution of ecosystem engineering, niche construction and diversity. Trends in Ecology and Evolution (23):304–10.

Faraji F., Eissenberg J.C. 2013. Seed and soil: A conceptual framework of metastasis for clinicians. Mo Med (110):302-8.

Fidler I.J. 1970. Metastasis: quantitative analysis of distribution and fate of tumor emboli labeled with 125 I-5-iodo-2'-deoxyuridine. J Natl Cancer Inst (45):773-82.

Fidler, I.J. 1973. Selection of successive tumor lines for metastasis. Nat New Biol (242):148-19.

Folkman J. 1971. Tumor angiogenesis: therapeutic implications. N Engl J Med (18): 1182-1186.

Fridman W.H., Pagés F., Satés-Fridman C., Galon J. 2012. The immune contexture in human tumours: impact on clinical outcome. Nat Rev Cancer (15):298-306.

Gatenby R.A. 1991. Population ecology issues in tumor growth. Cancer Res (51): 2542-2547.

Grinnell, J. 1917."The niche-relationships of the California Thrasher". Auk (34):427–433.

Gupta G.P., Massagué J. 2006. Cancer metastasis: building a framework. Cell (127): 679-695.

Hanahan D., Weinberg R.A. 2011. Hallmarks of cancer: the next generation. Cell 144(5):646-74

Healy, Michael, Mason, Travis V. and Ricou, Laurie. 2009. "'hardy/unkillable clichés': Exploring the Meanings of the Domestic Alien, Passer domesticus". Interdisciplinary Studies in Literature and Environment (Oxford University Press) 16 (2):281–298.

Hickman .J.E., Wu S., Mickley L.J., Lerdau M.T., Kudzu. 2010. (Pueraria montana) invasion doubles emissions of nitric oxide and increases ozone pollution. Proc Natl Acad Sci U S A 107(22):10115-9.

Hunter K. 2006. Host genetics influence tumour metastasis. Nat Rev Cancer (2):141-6.

Hutchinson, G.E. 1957. "Concluding remarks" Cold Spring Harbor Symposia on Quantitative Biology 22 (2):415–427.

Jain R.K. 2012. Delivery of molecular and cellular medicine to solid tumors. Adv Drug

Deliv Rev 64(Suppl):353-365.

Johnston, R. F., Selander, R. K. 1973. "Evolution in the House Sparrow. III. Variation in Size and Sexual Dimorphism in Europe and North and South America". The American Naturalist 107 (955):373–390.

Jones C. G., Lawton J. H., Shachak M. 1997. Positive and negative effects of organisms as physical ecosystem engineers. Ecology (78):1946–1957.

Jones C.G., Lawton J.H. and Shachak M. 1994. Organisms as ecosystem engineers. Oikos (69):373-386

Kim M.Y., Oskarsson T., Acharyya S., Nguyen D.X., Zhang X.H., Norton L., Massagué J. 2009. Tumor self-seeding by circulating cancer cells. Cell (139):1315-26

Klein C.A. 2013. Selection and adaptation during metastatic cancer progression. Nature 501(7467):365-72.

Knutson K.L., Disis M.L. 2005. Tumor antigen-specific T helper cells in cancer immunity and immunotherapy. Cancer Immunol Immunother (54):721-8.

Krakauer D. C., Page K. M., Erwin D. H. 2009. Diversity, dilemmas, and monopolies of niche construction. American Naturalist (173):26–40.

Kylafis, G., Loreau M. 2008. Ecological and evolutionary consequences of niche construction for its agent. Ecology Letters (11):1072–1081.

Lavi O., Greene J.M., Levy D., Gottesman M.M. 2014. Simplifying the complexity of resistance heterogeneity in metastasis. Trends Mol Med (3):129-136

Li W.A., Mooney D.J. 2013. Materials based tumor immunotherapy vaccines. Curr Opin Immunol (25):238–45.

Lianidou E.S., Mavroudis D., Pantel K. 2013. Advances in circulating tumor cells (ACTC): from basic research to clinical practice. Breast Cancer Res 15(6):319.

Loberg R.D., Bradley D.A., Tomlins SA, Chinnaiyan AM, Pienta KJ. 2007. The lethal phenotype of cancer: the molecular basis of death due to malignancy. CA Cancer J Clin 57(4):225-41

Loreau M. 2010. From Populations to Ecosystems: Theoretical Foundations for a New Ecological Synthesis. Princeton (New Jersey): Princeton University Press.

Lowes L.E., Allan A.L. 2014. Recent advances in the molecular characterization of circulating tumor cells. Cancers (Basel) 6(1):595-624

Maire V., Gross N., Börger L., Proulx R., Wirth C., da Silveira Pontes L., Soussana J.F., Louault F. 2012. Habitat filtering and niche differentiation jointly explain species relative abundance within grassland communities along fertility and disturbance gradients. New Phytol (2):497-509.

Matho L., Stenninger J. 2012. Behavior of seeds and soil in the mechanism of metastasis: A deeper understanding. Cancer Sci (103):626-631.

Mitchell M.J., Wayne E., Rana K., Schaffer C.B., King M.R. 2014. TRAIL-coated leukocytes that kill cancer cells in the circulation. Proc Natl Acad Sci USA (111):930-935.

Nguyen D.X., Bos P.D., Massagué J. 2009. Metastasis: from dissemination to organ-specific colonization. Nature Reviews Cancer (9):274-284.

Nishikawa M. 2008. Reactive oxygen species in tumor metastasis. Cancer Lett (18): 53-59.

Odling-Smee F.J., Erwin D.H., Palkovacs E.P, Feldman M.W., Laland K.N. 2013. Niche construction theory: a practical guide for ecologists. Q Rev Biol (1):4-28.

Odling-Smee F.J., Laland K.N., Feldman M.W. 2003. Niche Construction: The Neglected Process in Evolution. Princeton (New Jersey): Princeton University Press

Oskarsson T., Batlle E., Massagué J. 2014. Metastatic stem cells: sources, niches, and vital pathways. Cell Stem Cell (14):306-321

Paget S. 1989. The distribution of secondary growths in cancer of the breast. Lancet (1):571-573.

Pedersen E.A., Shiozawa Y., Pienta K.J., Taichman R.S. 2012. The prostate cancer bone marrow niche: more than just 'fertile soil'. Asian J Androl (14):423-427.

Pienta K.J., McGregor N., Axelrod R., Axelrod D.E. 2008. Ecological therapy for cancer: defining tumors using an ecosystem paradigm suggests new opportunities for novel cancer treatments. Transl Oncol (1):158-164.

Pienta K.J., Partin A.W., Coffey D.S. 1989. Cancer as a disease of DNA organization and dynamic cell structure. Cancer Res 49(10):2525-32

Pienta K.J., Robertson B.A., Coffey D.S., Taichman R.S. 2013. The cancer diaspora: Metastasis beyond the seed and soil hypothesis. Clin Cancer Res 19(21):5849-55.

Plaks V., Koopman C.D., Werb Z. 2013. Circulating Tumor Cells. Science (341):1186-1188.

Post D. M., Palkovacs E. P. 2009. Eco-evolutionary feedbacks in community and ecosystem ecology: interactions between the ecological theatre and the evolutionary play. Philosophical Transactions of the Royal Society B: Biological Sciences (364):1629–1640.

Ramakrishna R. and Rostomily R. 2013. Seed, soil, and beyond: The basic biology of brain metastasis. Surg Neurol Int (4):S256-S264.

Robertson B.A., Rehage J.S., Sih A. 2013. Ecological novelty and the emergence of evolutionary traps. Trends Ecol Evol (28):552–60.

Salerno E.P., Olson W.C., McSkimming C., Shea .S, lingluff C.L. Jr. 2014. T cells in the human metastatic melanoma microenvironment express site-specific homing receptors and retention integrins. Int J Cancer (134):563-574.

Scott J.G., Basanta D., Anderson A.R., Gerlee P. 2013. A mathematical model of tumour self-seeding reveals secondary metastatic deposits as drivers of primary tumour growth. J R Soc Interface (82):20130011

Scott J.G., Hjelmeland A.B., Chinnaiyan P., Anderson A.R., Basanta D. 2014. Microenvironmental variables must influence intrinsic phenotypic parameters of cancer stem cells to affect tumourigenicity. PLoS Comput Biol (1):e1003433.

Semenza G.L. 2012. Molecular mechanisms mediating metastasis of hypoxic breast cancer cells. Trends Mol Med (18):534–43.

Shibue T., Weinberg R.A. 2009. Integrin 1-focal adhesion kinase signaling directs the proliferation of metastatic cancer cells disseminated in the lungs. Proc Natl Acad Sci USA (106):10290-10295.

Shiozawa Y., Pedersen E.A., Havens A.M., Jung Y., Mishra A., Joseph J., et al. 2011. Human prostate cancer metastases target the hematopoietic stem cell niche to establish footholds in mouse bone marrow. J Clin Invest (121):1298–312.

Sreevalsan S., Safe S. 2013. Reactive oxygen species and colorectal cancer. Curr Colorectal Cancer Rep (9):350-357.

Steeg P.S. 2006. Tumor metastasis: mechanistic insights and clinical challenges. Nat Med (12):895-904.

Stylianopoulos T., Jain R.K. 2013. Combining two strategies to improve perfusion and drug delivery in solid tumors. Proc Natl Acad Sci U S A 110(46):18632-7

Suarez A.V., Holway D.A., Case T.J. 2001. Patterns of spread in biological invasions dominated by long-distance jump dispersal: Insights from Argentine ants. Proc Natl Acad Sci U S A 98(3):1095–1100.

Tarhini A.A., Edington H., Butterfield L.H., Lin Y., Shuai Y., Tawbi H., Sander C., Yin Y., Holtzman M., Johnson J., Rao U.N., Kirkwood J.M. 2014. Immune monitoring of the circulation and the tumor microenvironment in patients with regionally advanced melanoma receiving neoadjuvant ipilimumab. PLoS One 9(2):e87705.

Thuiller W., Münkemüller T., Lavergne S., Mouillot D., Mouquet N., Schiffers K., Gravel D. 2013. A road map for integrating eco-ecolutionary processes into biodiversity models. Ecol Lett (16):94-105.

Tinhofer I., Saki M., Niehr F., Keilholz U., Budach V. 2014. Cancer stem cell characteristics of circulating tumor cells. Int J Radiat Biol (just accepted)

Van der Sanden B., Appaix F., Berger .F, Selek L., Issartel J.P., Wion D. 2013. Translation of the ecological trap concept to glioma therapy: the cancer cell trap concept. Future Oncol (9):817–24.

Van Dyken J. D., Wade M. J. 2012. Origins of altruism diversity II: runaway coevolution of altruistic strategies via "reciprocal niche construction." Evolution (66):2498–2513.

Wagstaff L., Kolahgar G., Piddini E. 2013. Competitive cell interactions in cancer: a cellular tug of war. Trends Cell Biol (23):160-167.

Wan L., Pantel K., Kang Y. 2013. Tumor metastasis: moving new biological insights into the clinic. Nat Med (19):1450-1464.

Wang J., Loberg R., Taichman R.S. 2006. The pivotal role of CXCL12 (SDF-1)/CXCR4 axis in bone metastasis. Cancer Metastasis Rev (25):573–87.

Waris G., Ahsan H. 2006. Reactive oxygen species: role in the development of cancer and various chronic conditions. J Carcinog ( 5):14.

Weiss L., Bronk J., Pickren J.W., Lane W.W. 1981. Metastatic patterns and target organ arterial blood flow. Invasion Metastasis (1):126–35.

Wey J.S., Stoeltzing O., Ellis L.M. 2004. Vascular endothelial growth factor receptors: expression and function in solid tumors. Clin Adv Hematol Oncol (1):37-45.

Yu P., Fu Y.X. 2006. Tumor-infiltrating T lymphocytes: friends or foes? Lab Invest (86):231-245.

Zetter B.R. 1998. Angiogenesis and tumor metastasis. Annu Rev Med (49):407-424.

Zitvogel L., Apetoh L., Ghiringhelli F., André F., Tesniere A., Koemer G. 2008. The anticancer immune respone: indispensable for therapeutic success? J Clin Invest (118):1991-2001.

Figure 1:

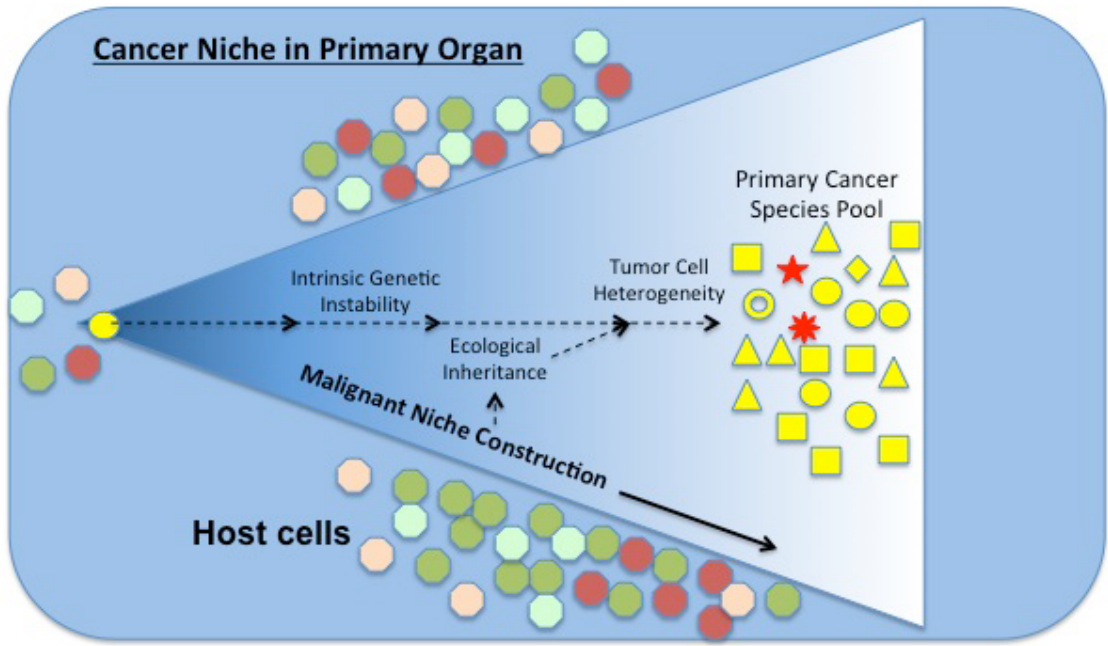

Figure 2

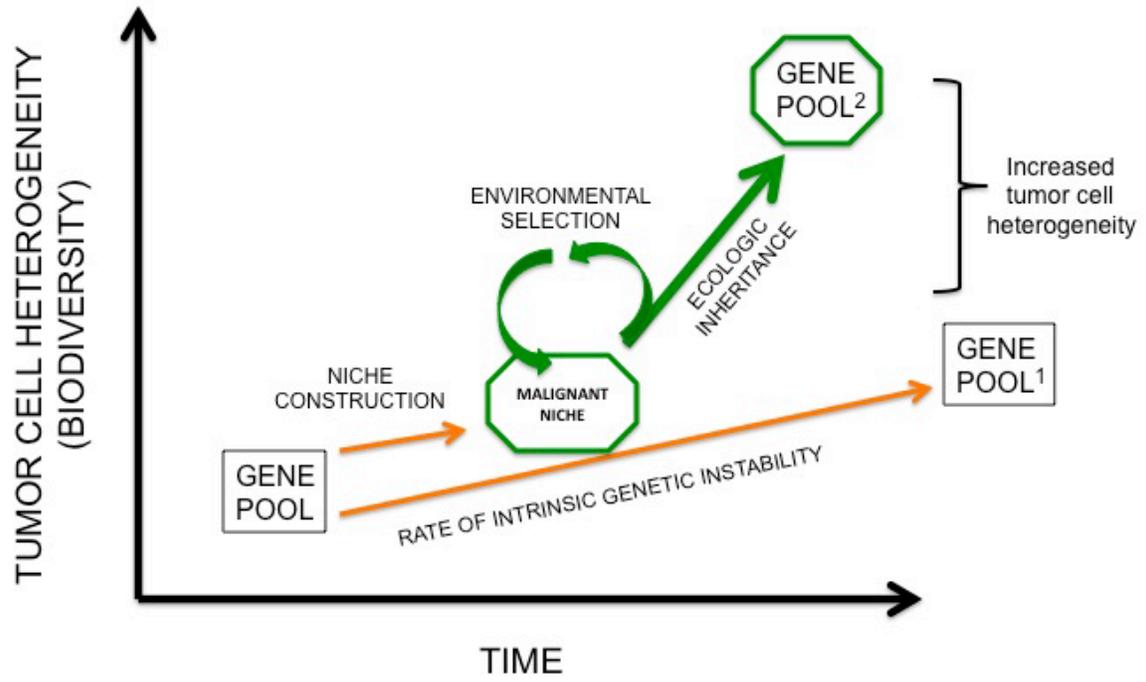

Figure 3:

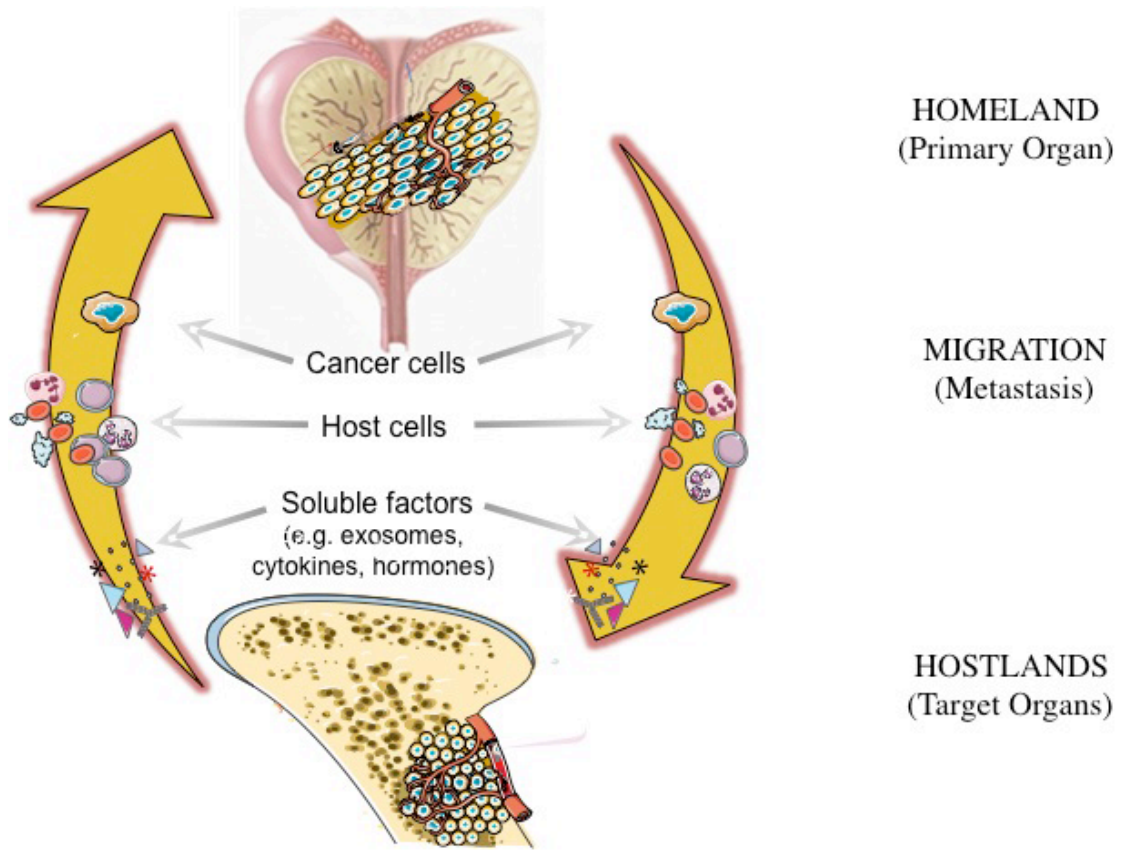

Figure 4:

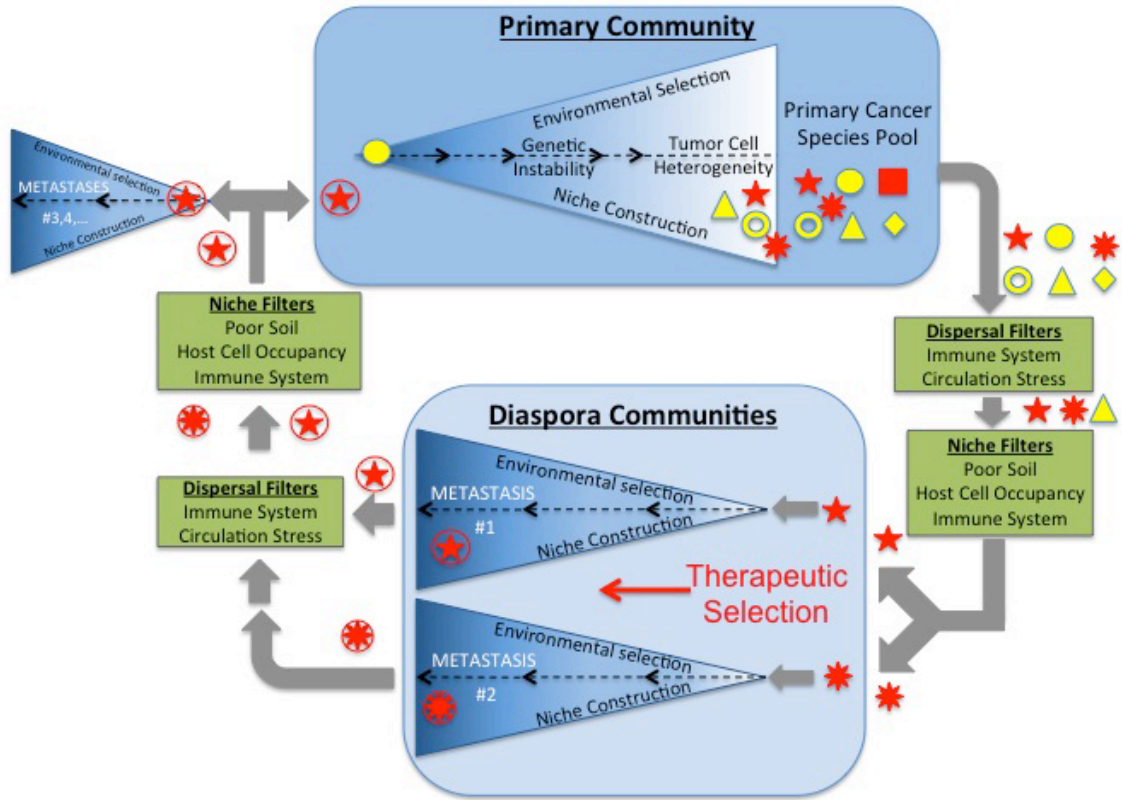